\documentclass{article}

\usepackage{arxiv}
\usepackage[utf8]{inputenc}
\usepackage[T1]{fontenc}
\usepackage{hyperref}
\usepackage{url}
\usepackage{doi}
\usepackage{graphicx}
\usepackage{booktabs}
\usepackage{longtable}
\usepackage{array}
\usepackage{tabularx}
\usepackage{ragged2e}
\usepackage{amsmath,amssymb}
\usepackage{enumitem}
\usepackage{microtype}
\usepackage{tikz}
\usetikzlibrary{arrows.meta,calc,positioning,shapes.geometric}
\usepackage[numbers]{natbib}
\hypersetup{%
    colorlinks=true,
    linkcolor=blue,
    citecolor=blue,
    urlcolor=blue,
    pdftitle={From Knowledge to Legitimacy: A Philosophical Problem Discovery of AI Implementation Readiness in Public Health Disease Surveillance},
    pdfauthor={Tithi Mitra, Souvik Pramanik, Most. Aysha Siddiqa Sumona, Ahmed Faizul Haque Dhrubo, Farhana Sharmin, Mohammad Abdul Qayum},
    pdfkeywords={Artificial Intelligence, Disease Surveillance, Implementation Readiness, Public Health, Digital Health Governance, Epistemic Adequacy, LMICs}
}

\newcolumntype{Y}{>{\RaggedRight\arraybackslash}X}
\AtBeginEnvironment{thebibliography}{\sloppy}

\title{\textbf{From Knowledge to Legitimacy: A Philosophical Problem Discovery of AI Implementation Readiness in Public Health Disease Surveillance}}

\author{
Tithi Mitra\\
Department of Philosophy\\
Jagannath University\\
Dhaka, Bangladesh\\
\texttt{m240106024@phil.jnu.ac.bd}
\And
Souvik Pramanik\\
Department of Electrical and Computer Engineering\\
North South University\\
Dhaka, Bangladesh\\
\texttt{souvik.pramanik@northsouth.edu}
\AND
Most. Aysha Siddika Sumona\\
Department of Electrical and Computer Engineering\\
North South University\\
Dhaka, Bangladesh\\
\texttt{ most.sumona@northsouth.edu}
\And
Ahmed Faizul Haque Dhrubo\thanks{Corresponding author: \texttt{ahmed.dhrubo@northsouth.edu}}\\
Department of Electrical and Computer Engineering\\
North South University\\
Dhaka, Bangladesh\\
\texttt{ahmed.dhrubo@northsouth.edu}
\AND
Farhana Sharmin\\
Department of Public Health\\
North South University\\
Dhaka, Bangladesh\\
\texttt{farhana.sharmin.242@northsouth.edu}
\And
Mohammad Abdul Qayum\\
Department of Electrical and Computer Engineering\\
North South University\\
Dhaka, Bangladesh\\
\texttt{mohammad.qayum@northsouth.edu}
}

\date{}
\renewcommand{\shorttitle}{AI Implementation Readiness in Disease Surveillance}

\begin{document}

\maketitle

\begin{abstract}

Despite being an emerging credible means of using artificial intelligence (AI) for improving disease surveillance via early detection of outbreaks, epidemics prediction and evidence-based decision-making, there continue to be challenges in the use of AI tools in many low- and middle-income countries (LMICs) due to various factors including fragmented health information system, digital inequality, governance problems and institutional incapability. Most of the research conducted so far has concentrated on the efficacy and accuracy of AI in terms of predicting outbreaks, with little focus on other conditions necessary for its deployment. This study adopts a qualitative problem-discovery research design, integrating thematic analysis with philosophical analysis to examine the structural and normative barriers surrounding AI implementation in disease surveillance. The analysis identifies four interrelated dimensions of implementation readiness: epistemic adequacy, distributive justice, ethics of governance, and institutional legitimacy. These dimensions provide a framework for understanding how limitations in knowledge integration, unequal digital infrastructure, privacy and accountability concerns, and deficits in institutional and public trust can constrain the practical adoption of AI-enabled surveillance. Rather than proposing another predictive model, this study develops a theoretical framework that conceptualizes AI implementation in disease surveillance as simultaneously a socio-technical and normative process. The framework provides a foundation for subsequent empirical investigation and offers a structured perspective for designing more context-sensitive, ethically grounded, and institutionally sustainable AI-enabled disease surveillance systems in resource-constrained healthcare settings.

\end{abstract}


\newpage
\section{Introduction}

The outbreak of infectious diseases has continued to demonstrate some of the persistent shortcomings of public health surveillance systems, especially in LMICs, which suffer from late reporting, lack of health data, poor digital connectivity, and limited institutional capacity to hamper prompt detection and response~\cite{ganser2022global}. Traditional surveillance systems still play an essential role in the practice of public health; however, their reliance on clinical diagnosis, laboratory analysis, and reporting systems at multiple levels may result in delays if the spread of a disease is fast-paced. These limitations have increased interest in approaches that can complement established surveillance mechanisms with more timely and heterogeneous sources of information.

The use of AI could potentially serve as a tool to improve the effectiveness of disease surveillance. The application of techniques from machine learning, natural language processing, and data fusion can aid in analyzing clinical, environmental, mobility, and internet data for identifying irregularity in disease occurrence, predicting its transmission dynamics, and directing resource allocation~cite{villanuevamiranda2025artificial,valentin2021monitoring}. However, the utilization of AI is not evenly distributed among health systems of LMICs~cite{ciecierskiholmes2022artificial}. Existing research has frequently emphasized the predictive performance and technical capabilities of AI systems, while comparatively less attention has been given to the conditions required for integrating such systems into routine public health practice~\cite{ahmed2023systematic}. The resulting gap is therefore not simply a question of whether AI can perform surveillance-related tasks, but whether the surrounding health system is sufficiently prepared to generate, govern, interpret, and act upon AI-enabled information.

This implementation challenge is inherently socio-technical. AI-enabled surveillance depends not only on computational capability but also on the quality and representativeness of the knowledge available to the system, the equitable distribution of the digital infrastructure required for its operation, the ethical legitimacy of health-data practices, and the institutional capacity and public trust necessary for sustained use. From the point of view of epistemology, restrictions on the production and representation of health information may restrict the kind of knowledge that can be attained by the AI, leading to doubts over the experiences and conditions that get depicted through data-based decision making~\cite{fricker2007epistemic}. From the point of view of ethics, the responsible use of AI requires taking into consideration aspects like beneficence, non-maleficence, autonomy, justice, and explicability~\cite{floridi2018ai}. The fragmented nature of health data may therefore restrict the kind of conclusions that the surveillance system can make; infrastructural inequality may restrict the reach of the benefits of AI-enabled capacities; weak privacy and accountability measures may weaken the legitimacy of data-based surveillance; and the lack of preparedness and trust may restrict the conversion of AI-generated information into public health practice. Therefore, implementation preparedness involves many other conditions apart from model performance.

Accordingly, this research takes a \emph{problem-discovery} approach rather than a research approach based on the proposal or evaluation of a new predictive model. Through an interdisciplinary review of existing literature in public health surveillance, artificial intelligence, digital governance, and health ethics, this leads to the discovery and systemization of the existing structural barriers to the implementation of AI-based disease surveillance in resource-limited public health systems. This study contributes to knowledge by (1) identifying four recurring implementation barriers, (2) structuring these barriers in the form of a conceptual framework, connecting technological, infrastructural, ethical, and institutional aspects, and (3) developing an agenda for further research on AI implementation readiness. By shifting the focus away from technological capacity and towards the wider conditions for implementation, this study creates the foundations for the analysis of how AI-based surveillance might be developed and implemented in a contextually appropriate manner.

% ==========================================================
% 2. BACKGROUND AND RELATED LITERATURE
% ==========================================================

\section{Background and Related Literature}

The use of AI technology in public health has seen a significant increase in the last decade in the field of surveillance of infectious diseases, prediction of outbreaks, epidemiological modeling, and resource allocation for health systems. The recent reviews indicate that the surveillance systems based on the use of AI technologies integrate heterogeneous sources of data, such as epidemiological data, data from the web, environmental and climate factors, mobility data, and others for the purpose of early detection and forecasting~\cite{villanuevamiranda2025artificial,santangelo2023machine}. Nonetheless, just having an ability to implement certain technologies is not equivalent to their implementation in practice. Recent studies reveal barriers related to various aspects, such as data quality, integration, infrastructure, interpretation, governance, workforce and adaptation to the context, especially in LMICs~\cite{ciecierskiholmes2022artificial,lopez2022challenges,alganad2026deploying}.

This knowledge gap needs to be understood against the background not only of the history of disease surveillance but also of the technological shift brought about by AI\@. In addition, it needs to be seen from the perspective of the circumstances under which AI-based knowledge becomes actionable public health knowledge. This chapter therefore presents a review of conventional surveillance systems, the evolution of AI-based approaches, their application and limitations, and relevant epistemological, ethical, distributive, and institutional issues.

% ----------------------------------------------------------

\subsection{Traditional Disease Surveillance}

Surveillance involves the continual monitoring and reporting of information concerning diseases in order to prevent and control diseases~\cite{choi2003understanding}. This is an essential element of public health that enables governments and healthcare organizations to track disease trends, discover new threats, conduct investigations on outbreaks and manage the health care resource allocation. Traditional surveillance is the basis upon which many modern AI systems have been developed.

Classic surveillance programs rely mainly on diagnosis confirmed in laboratories, clinical reporting, hospital reports, investigations into the epidemiology, mortality statistics, and national surveillance systems. They generate important clinical and epidemiological information, although their success is based on the timeliness of reporting and data collection at all levels of the health care system~\cite{ganser2022global}. The lack of interoperability and integration of the processes between different levels may hinder the flow of data from local providers of care to national surveillance programs.

These constraints are even more significant in constrained environments. Health institutions that exist in remote and underserved areas may still be using paper records or partially digitized systems, hence posing a problem in the timely transfer and aggregation of information~\cite{numair2021barriers}. In general, literature has noted the following as constraints to digitalizing the health system within LMICs - limited connectivity, incompatibility of information systems, incomplete data sets, lack of technological expertise, and lack of institutional support~\cite{ciecierskiholmes2022artificial,lopez2022challenges,alganad2026deploying}.

Traditional surveillance should therefore not be understood simply as an obsolete alternative to AI\@. Laboratory confirmation, clinical reporting, epidemiological investigation, and institutional notification remain essential components of public health surveillance. Rather, the limitation concerns the temporal and informational constraints of systems that depend primarily on confirmed clinical events and sequential reporting. These constraints have motivated the development of complementary approaches capable of incorporating heterogeneous and near-real-time information sources.

% ----------------------------------------------------------

\subsection{Artificial Intelligence in Public Health}

AI is generally defined as computational approaches that can detect patterns, learn from data, make predictions, and inform decisions. In public health, AI has broadened its scope beyond diagnostics in clinical care and into areas such as population surveillance, forecasting, outbreak detection, risk analysis, and health system management~\cite{ciecierskiholmes2022artificial}.

Machine learning and associated computing methods can detect non-linear associations present in large and diverse databases which are not possible to discover through traditional monitoring and statistics. AI-enabled systems can integrate electronic health records, laboratory data, environmental and climate variables, mobility patterns, and digital information streams to support the identification of abnormal disease activity and estimation of epidemiological risk~\cite{villanuevamiranda2025artificial,santangelo2023machine}. Recent systematic evidence on infectious disease early-warning systems confirms the growing use of machine learning, deep learning, and natural language processing across diverse epidemiological and digital data sources~\cite{villanuevamiranda2025artificial}.

Event-based digital surveillance provides one example of this transition. Natural language processing and machine learning can process large volumes of online news, reports, and other unstructured information to identify signals potentially associated with emerging outbreaks~\cite{valentin2021monitoring,villanuevamiranda2025artificial}. Similarly, predictive models have been applied to diseases such as COVID-19, influenza, and malaria using combinations of historical incidence, environmental, mobility, and other contextual variables~\cite{santangelo2023machine}.

AI has also been applied to operational public-health decision making. During the COVID-19 pandemic, predictive and simulation-based approaches were used to estimate hospital and intensive-care demand and to support resource planning~\cite{wikmanjorgensen2024hospitalization}. Such applications illustrate a broader transition from surveillance systems primarily concerned with recording confirmed events toward systems that also attempt to anticipate emerging risks and support forward-looking decisions.

Nonetheless, the switch does not negate the fundamental needs for public health surveillance. AI systems are still reliant on the accuracy, completeness, representativeness, and availability of information used to train the system and create predictions. As such, enhanced computing power does not necessarily address any shortcomings with the health information system used for surveillance.

% ----------------------------------------------------------

\subsection{Current State of AI-Based Disease Surveillance}

The current literature indicates three major areas in which AI contributes to disease surveillance and related public-health decision making.

Firstly, AI contributes to \textbf{early detection of an outbreak}. The technology can use natural language processing, machine learning, and event-based surveillance techniques to analyze information from various sources such as news on the web, reports and others and find any indicators of emerging infectious diseases~\cite{villanuevamiranda2025artificial,valentin2021monitoring}. It is proven that in most cases AI-powered early warning systems combine epidemiological, web, environmental, climatic, and wastewater data~\cite{villanuevamiranda2025artificial}.

Secondly, there is the aspect of how AI is used in \textbf{predictive epidemiological modeling}. Using machine learning algorithms, it would be possible to combine historical incidence figures with environmental, climatic, mobility, and demographic factors to forecast disease trends and transmission dynamics~\cite{santangelo2023machine}. In addition, a lot of work has shown that there is potential for using AI as an adjunct to infectious diseases modeling~\cite{getchell2026platforms}.

Thirdly, AI supports \textbf{health-system planning and resource allocation}. Predictive models have been applied to hospital and intensive-care demand forecasting, thereby supporting operational planning during periods of rapidly changing healthcare demand~\cite{wikmanjorgensen2024hospitalization}. Although these applications are not themselves disease-surveillance systems, they demonstrate how AI-generated forecasts can be connected to downstream public-health decision processes.

Recent platform-level research further indicates that AI-enabled infectious disease surveillance is developing into multiple system architectures, including early-warning networks, situational-awareness platforms, and integrated surveillance platforms~\cite{getchell2026platforms}. This development suggests that the field is moving beyond isolated predictive models toward more complex information infrastructures that combine multiple data sources and support public-health decision making.

The distribution of evidence is, however, uneven. The review of AI in the LMIC health systems shows limited evidence of AI use in the real world with issues of lack of data, workflow integration, user acceptability, context-specificity, connectivity, and costs~\cite{ciecierskiholmes2022artificial}. Recent reviews show that fragmented data systems, lack of infrastructure, local technical capacity and governance are major challenges to the sustainable implementation of AI in low resource settings~\cite{lopez2022challenges,alganad2026deploying}.

% ----------------------------------------------------------

\subsection{Beyond Predictive Performance: Implementation Readiness}

A central limitation of the existing literature is the tendency to evaluate AI primarily through technical performance measures. Accuracy, sensitivity, specificity, area under the receiver operating characteristic curve, forecasting error, and related metrics are important for evaluating predictive systems, but they do not by themselves establish whether a system can function effectively within a public-health institution.

Evidence from recent years demonstrates this gap. According to a systematic review of AI applications for healthcare-associated infection surveillance carried out in 2025, good model performance was typical; however, there is not much research on implementation in actual use and clinical settings, and only a few studies have investigated factors like workload, cost, or effects on patients~\cite{cozzolino2025surveillance}. In addition, systematic reviews of AI in LMIC health systems have found several barriers to implementation in practice, including workflow disruptions, user-unfriendly design, lack of contextualization, data deficiencies, and lack of effectiveness evidence~\cite{ciecierskiholmes2022artificial}.

This distinction can be expressed as:

\begin{equation}
\text{Technical Capability} \neq \text{Implementation Readiness}.
\end{equation}

Technical capability concerns what an AI system can achieve under specified experimental conditions. Implementation readiness concerns whether the surrounding health system possesses the informational, infrastructural, ethical, organizational, and human conditions necessary to use that capability reliably and responsibly.

This distinction is particularly important in LMIC settings. Recent reviews identify unreliable connectivity, fragmented information systems, limited technical skills, inadequate governance, and uneven resource distribution as interconnected barriers to AI deployment~\cite{ciecierskiholmes2022artificial,lopez2022challenges,alganad2026deploying}. Consequently, the implementation problem cannot be reduced to improving algorithms. It also requires strengthening the environment in which those algorithms operate.

% ----------------------------------------------------------

\subsection{Epistemic Dimensions of AI-Enabled Surveillance}

The implementation problem can also be interpreted from an epistemological perspective. Public-health surveillance is fundamentally a system for producing, validating, communicating, and acting upon knowledge about population health. AI does not directly observe disease activity; rather, it generates inferences from the data made available to it. The quality and representativeness of those data therefore establish important conditions for the knowledge that AI systems can produce.

Such a standpoint is useful when considering epistemic injustice, whereby one considers how societal or institutional factors can lead to the marginalization of certain people's knowledge, experience, or point of view when it comes to knowledge formation~\cite{fricker2007epistemic}. As far as the use of AI in surveillance is concerned, fragmented or systematically incomplete health data can restrict the visibility of certain groups of people, certain places, or certain disease outbreaks. It does not mean that data fragmentation is epistemic injustice per se; it shows that different information infrastructure can lead to epistemic barriers in public health decision making.

The epistemic question is therefore not simply whether an AI model is statistically accurate, but also whether the information environment provides an adequate basis for the claims generated by the system. Data completeness, representativeness, interoperability, provenance, and contextual validity consequently become part of the broader implementation-readiness problem.

% ----------------------------------------------------------

\subsection{Justice, Ethics, and Governance}

Another aspect related to the use of AI technology for surveillance involves distribution and ethical considerations that go far beyond just issues of effectiveness. Access to digital infrastructure, computing power, data, and skilled people is not evenly spread throughout the health care system. Therefore, the implementation of AI might be more feasible in the environment where there is already an advantage in terms of technological and organizational capabilities.

From the perspective of justice as fairness, this raises questions about whether the institutional conditions required to benefit from AI are equitably available, particularly to populations and regions that are already disadvantaged~\cite{rawls2001justice}. Contemporary work on AI deployment in low-resource settings similarly emphasizes the importance of addressing infrastructure inequality, connectivity, local capacity, and contextual adaptation rather than treating AI as a technology that can be transferred independently of its implementation environment~\cite{wong2025digital,alganad2026deploying}.

There emerges the additional issue of ethical governance. AI-based surveillance could use sensitive medical information and have consequences for people or communities even in cases when the technology is supposed to be used for public health purposes. The issues at hand are privacy, autonomy, data protection, accountability, transparency, fairness, and possible harms that come as a result of inappropriate outputs of the technology. The World Health Organization highlights human autonomy, safety, transparency, accountability, inclusivity, and equity as the key issues related to the use of AI in healthcare~\cite{who2021ethics}. Floridi et al. define responsible AI based on such principles as beneficence, non-maleficence, autonomy, justice, and explicability~\cite{floridi2018ai}.

These perspectives imply that governance cannot be seen as an additional regulatory overlay placed on top of the technological innovations developed. Instead, governance from the point of ethics and institutions is an integral part of the context within which AI-driven surveillance can be rendered legitimate and sustainable. Issues of privacy protection, explainability, accountability, and well-defined institutional roles may affect the ability of relevant actors to trust AI-generated data.

% ----------------------------------------------------------

\subsection{Limitations of Existing Literature}

All in all, the literature clearly indicates a lot of advancement in the technical aspects of AI-powered public health surveillance, as well as the existence of certain implementation challenges. The current literature has produced information on early detection, epidemiological prediction, and forecasting; however, most of the available literature focuses on the performance of algorithms in ideal conditions~\cite{villanuevamiranda2025artificial,santangelo2023machine}. Information on implementation, institutionalization, and health benefits is rather scarce~\cite{cozzolino2025surveillance,ciecierskiholmes2022artificial}.

The second limitation pertains to the division of implementation challenges across various research paradigms. Data quality and data interoperability are addressed as technical challenges; infrastructure and connectivity as health system challenges; privacy and accountability as ethical or regulatory challenges; and trust and workforce as social challenges. Despite these distinctions, recent reviews have found these issues to be interrelated challenges~\cite{lopez2022challenges,alganad2026deploying}.

This fragmentation creates a conceptual gap. The literature provides substantial evidence that individual implementation barriers exist, but less attention has been devoted to organizing these barriers into a unified account of \emph{AI implementation readiness} for low-resource public-health surveillance. In particular, the relationship between the epistemic foundations of surveillance, the equitable distribution of digital capacity, the ethical legitimacy of data practices, and institutional trust remains insufficiently integrated.

This research fills the existing gap by taking a problem discovery perspective. It does not propose any additional predictive models but draws upon cross-disciplinary findings in order to discover structural problems and interpret them from technological, infrastructural, epistemic, ethical, and institutional perspectives. It allows developing a theoretical framework proposed later in the paper. Table~\ref{tab:literature_review} below gives an overview of the representative literature used in the synthesis.

% =============================================================================
% FIGURE: From Technical Capability to AI Implementation Readiness
% =============================================================================

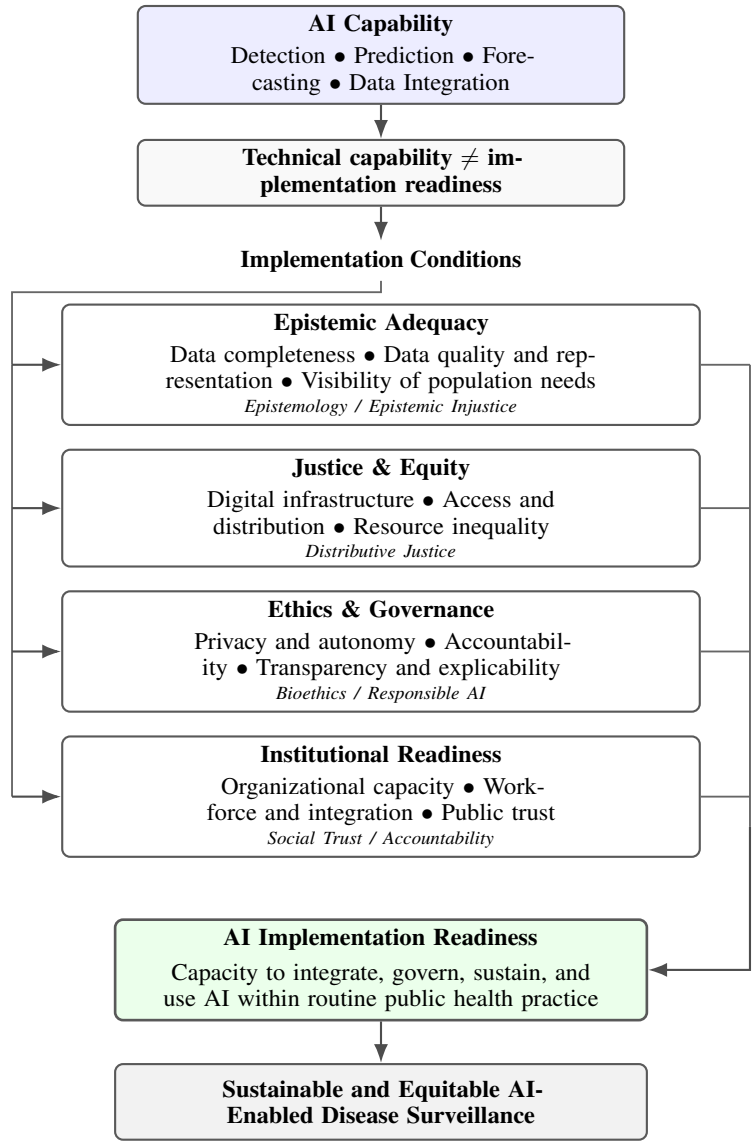
\begin{figure}[p]
\centering

\begin{tikzpicture}[
    font=\small,
    >=Latex,
    mainbox/.style={
        rectangle,
        rounded corners=3pt,
        draw=black!65,
        line width=0.8pt,
        align=center
    },
    capability/.style={
        mainbox,
        text width=6.2cm,
        minimum height=1.15cm,
        fill=blue!7
    },
    distinction/.style={
        mainbox,
        text width=6.2cm,
        minimum height=0.65cm,
        fill=gray!5,
        font=\bfseries\footnotesize
    },
    dimension/.style={
        mainbox,
        text width=8.2cm,
        minimum height=1.35cm,
        fill=white
    },
    readiness/.style={
        mainbox,
        text width=6.8cm,
        minimum height=1.2cm,
        fill=green!8,
        line width=1pt
    },
    outcome/.style={
        mainbox,
        text width=6.8cm,
        minimum height=1cm,
        fill=gray!10
    },
    arrow/.style={
        -{Latex[length=2.5mm]},
        line width=0.8pt,
        draw=black!70
    },
    connector/.style={
        line width=0.7pt,
        draw=black!60
    }
]

\node[capability] (capability) {
    \textbf{AI Capability}\\[2pt]
    \footnotesize Detection $\bullet$ Prediction $\bullet$ Forecasting $\bullet$ Data Integration
};

\node[distinction, below=0.45cm of capability] (distinction) {
    Technical capability $\neq$ implementation readiness
};

\node[font=\bfseries\footnotesize, below=0.45cm of distinction]
    (conditionslabel) {Implementation Conditions};

\node[dimension, below=0.3cm of conditionslabel] (epistemic) {
    \textbf{Epistemic Adequacy}\\[2pt]
    \footnotesize Data completeness $\bullet$ Data quality and representation $\bullet$ Visibility of population needs\\[-1pt]
    \scriptsize\itshape Epistemology / Epistemic Injustice
};

\node[dimension, below=0.3cm of epistemic] (justice) {
    \textbf{Justice \& Equity}\\[2pt]
    \footnotesize Digital infrastructure $\bullet$ Access and distribution $\bullet$ Resource inequality\\[-1pt]
    \scriptsize\itshape Distributive Justice
};

\node[dimension, below=0.3cm of justice] (ethics) {
    \textbf{Ethics \& Governance}\\[2pt]
    \footnotesize Privacy and autonomy $\bullet$ Accountability $\bullet$ Transparency and explicability\\[-1pt]
    \scriptsize\itshape Bioethics / Responsible AI
};

\node[dimension, below=0.3cm of ethics] (institutional) {
    \textbf{Institutional Readiness}\\[2pt]
    \footnotesize Organizational capacity $\bullet$ Workforce and integration $\bullet$ Public trust\\[-1pt]
    \scriptsize\itshape Social Trust / Accountability
};

\node[readiness, below=0.8cm of institutional] (readiness) {
    \textbf{AI Implementation Readiness}\\[3pt]
    \footnotesize Capacity to integrate, govern, sustain, and use AI within routine public health practice
};

\node[outcome, below=0.55cm of readiness] (outcome) {
    \textbf{Sustainable and Equitable AI-Enabled Disease Surveillance}
};

\draw[arrow] (capability) -- (distinction);
\draw[arrow] (distinction) -- (conditionslabel);
\draw[arrow] (readiness) -- (outcome);

\coordinate (condition-bus-top) at ($(epistemic.west)+(-0.65cm,0)$);
\coordinate (condition-bus-bottom) at ($(institutional.west)+(-0.65cm,0)$);
\draw[connector] (conditionslabel.south) -- ++(0,-0.15cm) -| (condition-bus-top);
\draw[connector] (condition-bus-top) -- (condition-bus-bottom);
\draw[arrow] (condition-bus-top) -- (epistemic.west);
\draw[arrow] (condition-bus-top |- justice.west) -- (justice.west);
\draw[arrow] (condition-bus-top |- ethics.west) -- (ethics.west);
\draw[arrow] (condition-bus-bottom) -- (institutional.west);

\coordinate (bus-top) at ($(epistemic.east)+(0.65cm,0)$);
\coordinate (bus-bottom) at ($(institutional.east)+(0.65cm,-0.4cm)$);
\draw[connector] (epistemic.east) -- (bus-top);
\draw[connector] (justice.east) -- (bus-top |- justice.east);
\draw[connector] (ethics.east) -- (bus-top |- ethics.east);
\draw[connector] (institutional.east) -- (bus-top |- institutional.east);
\draw[connector] (bus-top) -- (bus-bottom);
\draw[arrow] (bus-bottom) |- (readiness.east);

\end{tikzpicture}

\caption{Conceptual framework linking AI capability to implementation readiness in
low-resource public health systems. AI capability provides the technical basis for
disease surveillance, but practical implementation depends on epistemic adequacy,
justice and equity, ethical governance, and institutional readiness. These dimensions
converge to shape the conditions under which AI can be integrated into routine public
health practice and sustained in an equitable manner.}\label{fig:implementation_readiness_framework}

\end{figure}

The preceding literature suggests that AI capability alone does not determine
whether an AI-enabled surveillance system can be implemented in practice.
Accordingly, this study conceptualizes implementation readiness as a
socio-technical construct shaped by four interrelated dimensions: epistemic
adequacy, justice and equity, ethical governance, and institutional readiness.
Figure~\ref{fig:implementation_readiness_framework} summarizes this framework.

% ==========================================================
% Table: Representative literature
% ==========================================================

\begin{table}[t]
\centering
\scriptsize
\caption{Representative literature on AI-enabled public health surveillance and implementation in low-resource settings.}\label{tab:literature_review}
\begin{tabularx}{\textwidth}{@{}YYYY@{}}
\toprule

\textbf{Study} &
\textbf{AI Application / Focus} &
\textbf{Public Health Domain} &
\textbf{Key Limitation or Implementation Concern} \\
\midrule

Villanueva-Miranda et al.~\cite{villanuevamiranda2025artificial} &
Early-warning systems using ML, DL, and NLP &
Infectious disease surveillance &
Data quality and bias; transparency; system integration; privacy and equity concerns \\

Santangelo et al.~\cite{santangelo2023machine} &
Machine-learning prediction across multiple diseases &
Epidemiology and disease forecasting &
Heterogeneous evidence and limited external validation across settings \\

Wikman-Jorgensen et al.~\cite{wikmanjorgensen2024hospitalization} &
Hospital and ICU occupancy forecasting &
Health-system resource planning &
Limited disease-surveillance specificity and contextual validation \\

Ciecierski-Holmes et al.~\cite{ciecierskiholmes2022artificial} &
AI applications across diagnosis, triage, decision support, and health systems &
LMIC health systems &
Limited real-world evidence; data scarcity; workflow integration; trust and local-context challenges \\

López et al.~\cite{lopez2022challenges} &
Review of AI implementation requirements and barriers &
LMIC health ecosystems &
Data quality, infrastructure, connectivity, governance, workforce, trust, equity, and accountability barriers \\

Bostan et al.~\cite{bostan2024contextual} &
Contextual and infrastructural barriers to digital health systems &
LMIC electronic health systems &
Broad barrier coverage; relative severity and interactions remain insufficiently established \\

Wong et al.~\cite{wong2025digital} &
Implementation strategies for medical AI in low-resource settings &
Digital health and AI deployment &
Persistent infrastructure, workforce, and contextual-adaptation requirements \\

Cozzolino et al.~\cite{cozzolino2025surveillance} &
AI-based surveillance for healthcare-associated infections &
Healthcare-associated infection surveillance &
High reported model performance but limited evidence of real-world deployment and impact \\

Getchell et al.~\cite{getchell2026platforms} &
AI-enabled surveillance platform architectures &
Infectious disease surveillance &
Rapidly evolving platforms; implementation depends on data integration, localization, and governance \\

Al-Ganad et al.~\cite{alganad2026deploying} &
Scoping review of deployment barriers and strategies &
AI in low-resource settings &
Fragmented data, infrastructure, skills, governance, sustainability, and contextual adaptation remain persistent barriers \\

WHO~\cite{who2021ethics} &
Ethical and governance principles for AI in health &
Global health and public-health applications &
Requires attention to autonomy, safety, transparency, accountability, equity, and responsible governance \\

Floridi et al.~\cite{floridi2018ai} &
Ethical framework emphasizing beneficence, non-maleficence, autonomy, justice, and explicability &
AI ethics and governance &
Primarily normative framework; requires contextualization for public-health surveillance and LMIC implementation \\

\bottomrule
\end{tabularx}
\end{table}

% ==========================================================
% 3. RESEARCH PROBLEM
% ==========================================================

\section{Research Problem}

\subsection{Problem Statement}

The unresolved research problem is the absence of an integrated account of the conditions that determine whether AI-enabled disease surveillance can be implemented responsibly and effectively in resource-constrained public health systems. Existing research identifies individual technical, infrastructural, ethical, and organizational difficulties~\cite{ciecierskiholmes2022artificial,ahmed2023systematic}, yet these challenges are often examined in isolation.

As a result, current literature provides limited understanding of how the epistemic foundations of surveillance, the equitable distribution of digital infrastructure, the ethical governance of health data, and the institutional conditions necessary for public trust collectively shape implementation readiness. Consequently, it remains unclear how AI-generated knowledge can become sufficiently reliable, legitimate, and actionable within routine public health practice, particularly in low- and middle-income countries (LMICs).

This study therefore asks:

\textbf{\emph{What recurring barriers constrain the implementation of AI-enabled disease surveillance in low-resource public health systems, and how can these barriers be understood through epistemic, ethical, distributive, and institutional dimensions of implementation readiness?}}

% ----------------------------------------------------------
\subsection{Research Objectives}

The primary objective of this study is to identify and conceptualize the conditions that shape implementation readiness for AI-enabled disease surveillance in resource-constrained public health systems.

The specific objectives are:

\begin{enumerate}

    \item To examine the structural limitations of traditional disease surveillance that motivate the adoption of AI-enabled approaches.

    \item To identify the major barriers affecting AI implementation, including challenges related to data infrastructure, digital inequality, governance, and institutional capacity.

    \item To analyze these barriers through epistemic, ethical, distributive, and institutional perspectives in order to understand how they influence the production, legitimacy, and use of AI-generated public health knowledge.

    \item To develop a conceptual framework explaining how these dimensions collectively shape AI implementation readiness within LMIC public health systems.

    \item To establish a research agenda that can guide future empirical investigations into AI implementation, digital health governance, and equitable public health innovation.

\end{enumerate}

% ==========================================================
% 4. METHODOLOGY
% ==========================================================

\section{Methodology}

\subsection{Research Design}

This study adopts a qualitative \textit{problem discovery} research design using an interdisciplinary methodology that combines thematic synthesis with philosophical conceptual analysis. Rather than developing or evaluating an artificial intelligence algorithm, the study investigates the conditions that determine whether AI-enabled disease surveillance can be implemented responsibly within resource-constrained public health systems.

The research treats implementation readiness as a socio-technical concept situated at the intersection of public health, artificial intelligence, health informatics, digital governance, epistemology, and ethics. Accordingly, the study seeks not only to identify recurring implementation barriers, but also to interpret their epistemic, distributive, ethical, and institutional significance.

\subsection{Data Sources}

The analysis is based exclusively on secondary sources, including peer-reviewed journal articles, systematic reviews, international public health guidelines, and authoritative policy reports. Literature was selected from interdisciplinary domains including infectious disease surveillance, artificial intelligence, digital health, health governance, bioethics, and philosophy of knowledge.

Representative sources include publications from the World Health Organization (WHO), Centers for Disease Control and Prevention (CDC), \textit{The Lancet}, \textit{Nature}, \textit{PLOS}, \textit{npj Digital Medicine}, and other indexed public health and digital health journals.

\subsection{Selection Criteria}

The literature was selected through purposive sampling to ensure relevance to the research problem. Sources were included when they satisfied one or more of the following criteria:

\begin{itemize}
    \item examined AI applications in public health or disease surveillance;
    \item investigated infectious disease surveillance, outbreak prediction, or digital epidemiology;
    \item discussed implementation barriers involving infrastructure, governance, ethics, privacy, or institutional capacity;
    \item addressed low- and middle-income countries (LMICs) or other resource-constrained healthcare systems; or
    \item provided philosophical or normative frameworks relevant to epistemology, justice, or responsible AI governance.
\end{itemize}

Publications concerned exclusively with clinical diagnosis or algorithm development without population-level surveillance or implementation relevance were excluded.

\subsection{Analytical Strategy}

The study employs a two-stage analytical strategy consisting of thematic synthesis and philosophical conceptual analysis. Thematic synthesis follows the approach of Thomas and Harden~\cite{thomasharden2008methods} to identify recurring implementation barriers across interdisciplinary literature, while conceptual analysis is used to interpret the normative meaning of those themes through established philosophical frameworks.

The analysis proceeded through six stages:

\begin{enumerate}
    \item Literature collection
    \item Initial screening
    \item Open coding of implementation challenges
    \item Theme development
    \item Philosophical interpretation of themes
    \item Construction of the conceptual framework
\end{enumerate}

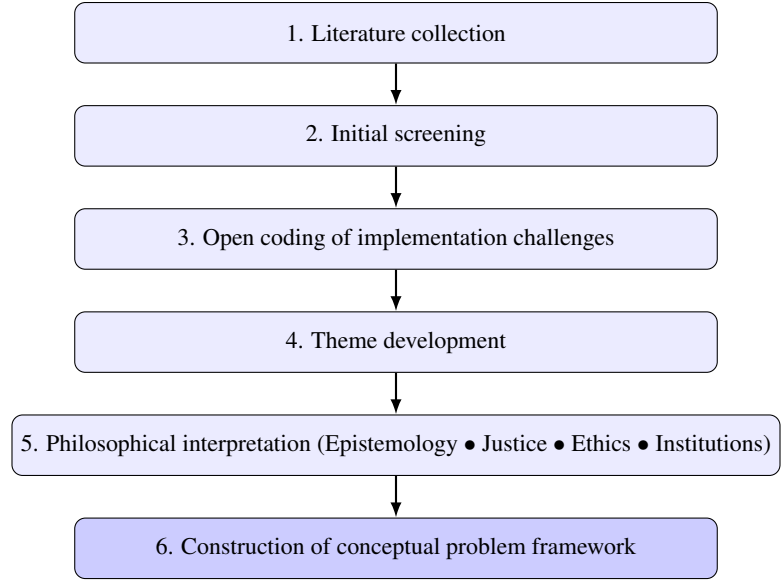
\begin{figure}[t]
\centering
\begin{tikzpicture}[
    node distance=0.55cm,
    stage/.style={rectangle, draw, rounded corners, fill=blue!8, minimum width=8.5cm, minimum height=0.8cm, align=center, font=\small},
    arr/.style={-{Latex[length=2mm]}, thick}
]
    \node[stage] (s1) {1. Literature collection};
    \node[stage, below=of s1] (s2) {2. Initial screening};
    \node[stage, below=of s2] (s3) {3. Open coding of implementation challenges};
    \node[stage, below=of s3] (s4) {4. Theme development};
    \node[stage, below=of s4] (s5) {5. Philosophical interpretation (Epistemology $\bullet$ Justice $\bullet$ Ethics $\bullet$ Institutions)};
    \node[stage, below=of s5, fill=blue!20] (s6) {6. Construction of conceptual problem framework};

    \draw[arr] (s1) -- (s2);
    \draw[arr] (s2) -- (s3);
    \draw[arr] (s3) -- (s4);
    \draw[arr] (s4) -- (s5);
    \draw[arr] (s5) -- (s6);
\end{tikzpicture}
\caption{Workflow of the qualitative problem discovery methodology, adapted from the six-phase thematic analysis process of Braun and Clarke~\cite{braunclarke2006using}.}\label{fig:methodology}
\end{figure}

During the thematic stage, implementation challenges were coded according to recurring patterns relating to data infrastructure, digital capacity, governance, privacy, institutional readiness, and public trust. In the second stage, these themes were interpreted through four philosophical dimensions identified in the literature: epistemic adequacy, distributive justice, ethical governance, and institutional legitimacy. Rather than testing causal relationships statistically, the analysis explains how these interrelated dimensions collectively shape AI implementation readiness in public health systems. Figure~\ref{fig:methodology} presents the six-stage analytical workflow, while Table~\ref{tab:coding} summarizes the relationship between the thematic codes and their philosophical interpretation.

% ----------------------------------------------------------
% Figure 2: Methodology workflow (TikZ)
% ----------------------------------------------------------

% ----------------------------------------------------------
% Table 2: Thematic coding framework (filled)
% ----------------------------------------------------------

\begin{table}[t]
\centering
\caption{Thematic coding framework and philosophical interpretation.}\label{tab:coding}
\begin{tabularx}{\textwidth}{@{}>{\RaggedRight\arraybackslash}p{3.4cm}Y>{\RaggedRight\arraybackslash}p{4.2cm}@{}}
\toprule
\textbf{Code} & \textbf{Implementation Theme} & \textbf{Philosophical Dimension}\\
\midrule
Data Infrastructure & Fragmentation, interoperability, and data quality & Epistemic adequacy\\

Digital Capacity & Connectivity, hardware, cloud infrastructure & Justice and equity\\

Governance & Regulation, accountability, oversight & Ethical governance\\

Privacy & Consent, confidentiality, data protection & Ethical governance\\

Institutional Readiness & Workforce, organizational capacity, integration & Institutional legitimacy\\

Public Trust & Citizen and clinician confidence in AI & Institutional legitimacy\\
\bottomrule
\end{tabularx}
\end{table}

% ==========================================================
% 5. PROBLEM DISCOVERY
% ==========================================================

\section{Problem Discovery}

The thematic synthesis identified four interrelated implementation barriers that consistently constrain the adoption of AI-enabled disease surveillance in resource-constrained public health systems. Rather than treating these barriers as isolated technical failures, the analysis interprets them as socio-technical conditions that shape AI implementation readiness. Through philosophical conceptual analysis, each barrier corresponds to a broader dimension of implementation: epistemic adequacy, distributive justice, ethical governance, and institutional legitimacy.

%----------------------------------------------------------

\subsection{Epistemic Barrier: Fragmented Health Data Infrastructure}

Disease surveillance is fundamentally a knowledge-producing activity. AI systems do not directly observe disease outbreaks; they generate inferences from the health information available to them. Consequently, the quality, completeness, and interoperability of health data determine the epistemic adequacy of AI-generated knowledge.

Across many LMICs, health information remains fragmented between hospitals, laboratories, regional health offices, and paper-based administrative records~\cite{numair2021barriers}. The absence of interoperable electronic health records produces incomplete datasets that reduce both the reliability and scalability of machine-learning models~\cite{ciecierskiholmes2022artificial}.

This fragmentation creates three interconnected consequences. First, disease reporting becomes delayed because information must pass through multiple administrative levels before reaching surveillance authorities~\cite{ganser2022global}. Second, incompatible data standards limit information sharing across healthcare institutions. For example, Bangladesh's attempted interoperability between the e-TB Manager and DHIS2 platforms illustrates how independently developed digital systems can remain structurally disconnected despite technological availability~\cite{siaps2017dhis2}. Third, incomplete or inaccurate records reduce the validity of AI predictions regardless of algorithmic sophistication.

From an epistemological perspective, fragmented data do not merely weaken technical performance; they constrain what AI systems are capable of knowing about population health. Implementation readiness therefore begins with the production of reliable and representative public-health knowledge rather than computational capability alone.

%----------------------------------------------------------

\subsection{Distributive Barrier: Digital Infrastructure Inequality}

Even when AI technologies are technically effective, their benefits depend upon the equitable distribution of digital infrastructure. Reliable internet connectivity, cloud computing, secure databases, and adequately equipped healthcare facilities constitute essential preconditions for AI deployment~\cite{numair2021barriers}.

The literature consistently demonstrates that these resources are distributed unevenly. A scoping review of electronic health record implementation in LMICs found that unreliable electricity, limited connectivity, and organizational resource constraints remain persistent barriers outside well-resourced urban centres~\cite{bostan2024contextual}. Likewise, documented AI deployments continue to be concentrated in institutions that already possess stronger digital infrastructure, while comparatively few reach geographically remote or underserved populations~\cite{ciecierskiholmes2022artificial}.

This inequality represents more than a technological gap; it raises a question of distributive justice. If access to AI-enabled surveillance depends upon existing institutional privilege, then the populations most vulnerable to infectious disease may simultaneously become those least able to benefit from technological innovation. Equitable implementation therefore requires reducing infrastructural disparities rather than simply expanding AI capability.

%----------------------------------------------------------

\subsection{Ethical Barrier: Privacy and Governance}

AI-enabled surveillance requires the collection and analysis of sensitive health information, including demographic characteristics, medical histories, laboratory records, and mobility data. While these datasets strengthen predictive surveillance, they simultaneously generate concerns regarding privacy, informed consent, transparency, accountability, and responsible data governance~\cite{ahmed2023systematic}.

These challenges are particularly significant within LMICs, where legal and institutional mechanisms governing health data often remain unevenly developed~\cite{tiffin2019howtouse}. Existing literature frequently recognizes privacy as an ethical principle, yet comparatively less attention is devoted to practical governance questions involving data ownership, cross-institutional sharing, algorithmic accountability, and explainability~\cite{ahmed2023systematic,tiffin2019howtouse}.

The philosophical significance of this barrier lies in distinguishing ethical principles from institutional governance. Responsible AI is not achieved solely by declaring values such as autonomy or beneficence; it also requires regulatory structures capable of enforcing transparency, protecting confidentiality, and assigning accountability for AI-supported public-health decisions. Ethical governance therefore functions as a necessary condition for legitimate implementation rather than an external regulatory addition.

%----------------------------------------------------------

\subsection{Institutional Barrier: Readiness and Public Trust}

The final barrier concerns the institutional and social conditions under which AI-generated knowledge becomes actionable. Successful implementation depends not only on accurate prediction but also on organizational capacity, workforce preparedness, regulatory coordination, and public trust.

Evidence from clinician-focused studies shows that healthcare professionals are substantially less willing to adopt AI tools when they perceive them as increasing workload, introducing clinical risk, or producing opaque recommendations~\cite{shamszare2023clinicians}. Similar concerns appear within LMIC deployments, where limited transparency regarding training data and black-box decision making has contributed to clinician distrust and reduced adoption~\cite{ciecierskiholmes2022artificial}.

Public trust represents a complementary institutional challenge. Citizens are less likely to participate in digital surveillance initiatives when privacy protections and governmental accountability appear uncertain. The experience of Singapore's TraceTogether programme demonstrates that sustained participation depended as much upon institutional trust and perceived legitimacy as upon technological effectiveness~\cite{kang2024roles}.

Institutional readiness should therefore be understood as a condition of legitimacy. AI implementation succeeds only when healthcare organizations possess the capacity to integrate AI into routine workflows and when both clinicians and the public regard AI-supported surveillance as trustworthy, accountable, and socially legitimate. Table~\ref{tab:problem_matrix} consolidates the four barriers, their corresponding philosophical dimensions, implementation impacts, and relative priorities.

%----------------------------------------------------------
% Table 3: Problem Matrix
%----------------------------------------------------------

\begin{table}[t]
\centering
\caption{Problem matrix linking implementation barriers with philosophical dimensions.}\label{tab:problem_matrix}

\begin{tabularx}{\textwidth}{@{}>{\RaggedRight\arraybackslash}p{2.7cm}>{\RaggedRight\arraybackslash}p{2.8cm}Y>{\RaggedRight\arraybackslash}p{1.5cm}@{}}
\toprule
\textbf{Barrier} & \textbf{Philosophical Dimension} & \textbf{Implementation Impact} & \textbf{Priority}\\
\midrule

Fragmented health data & Epistemic adequacy & Produces incomplete knowledge, delays outbreak reporting, and reduces the reliability of AI predictions. & High\\

Digital infrastructure inequality & Distributive justice & Unequal connectivity and digital resources concentrate AI benefits in better-resourced institutions. & High\\

Privacy and governance & Ethical governance & Weak accountability and data protection undermine responsible information sharing and regulatory legitimacy. & Medium\\

Institutional readiness and public trust & Institutional legitimacy & Limited organizational capacity and distrust reduce clinician adoption, public participation, and long-term sustainability. & High\\

\bottomrule
\end{tabularx}

\end{table}

% ==========================================================
% 6. CONCEPTUAL FRAMEWORK
% ==========================================================

\section{Conceptual Framework}

The findings suggest that AI implementation readiness is not determined by algorithmic performance alone, but by the interaction between technological capability and the socio-technical conditions within which AI operates. The thematic and philosophical analyses indicate that successful disease surveillance depends upon four interdependent dimensions: epistemic adequacy, distributive justice, ethical governance, and institutional legitimacy.

These dimensions do not function independently. Epistemic adequacy determines whether surveillance systems generate reliable and representative public-health knowledge; distributive justice shapes whether digital infrastructure and AI capabilities are equitably accessible; ethical governance establishes the legitimacy of data collection, privacy protection, and algorithmic accountability; and institutional legitimacy enables healthcare professionals and the public to trust and integrate AI into routine practice.

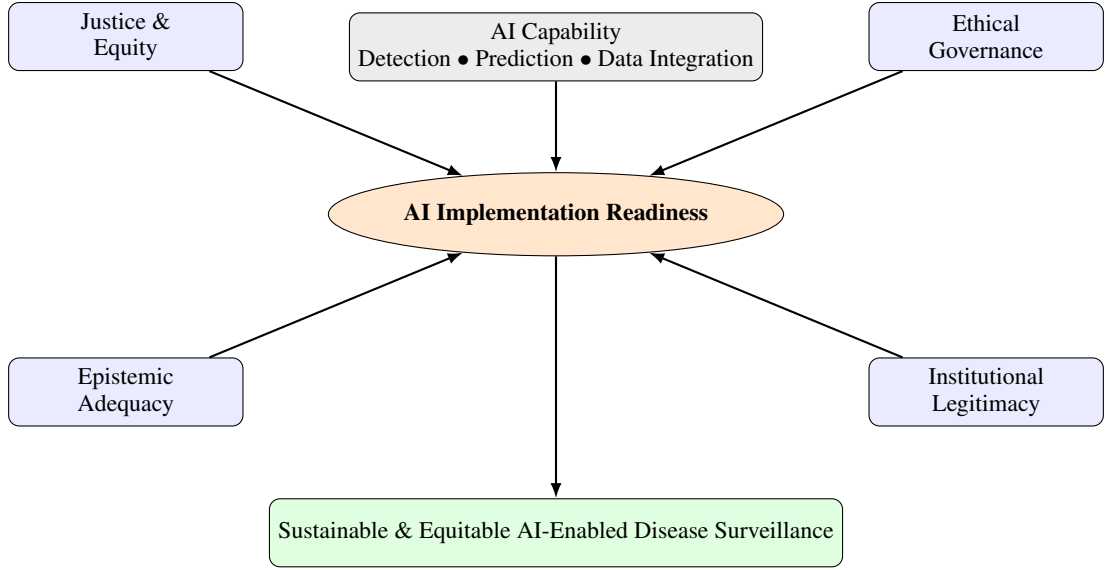
\begin{figure}[t]
\centering
\begin{tikzpicture}[
    node distance=1.2cm and 1.5cm,
    stage/.style={rectangle, draw, rounded corners, align=center, minimum width=3.1cm, minimum height=0.9cm, font=\small, fill=blue!8},
    core/.style={ellipse, draw, align=center, minimum width=4.1cm, minimum height=1.1cm, font=\small\bfseries, fill=orange!20},
    outcome/.style={rectangle, draw, rounded corners, align=center, minimum width=5cm, minimum height=0.9cm, font=\small, fill=green!12},
    arr/.style={-{Latex[length=2mm]}, thick}
]

% Top
\node[stage, fill=gray!15] (cap) {AI Capability\\Detection $\bullet$ Prediction $\bullet$ Data Integration};

% Middle
\node[core, below=of cap] (ready) {AI Implementation Readiness};

% Four dimensions
\node[stage, below left=1.5cm and 2cm of ready] (e1) {Epistemic\\Adequacy};

\node[stage, above left=1.5cm and 2cm of ready] (e2) {Justice \&\\Equity};

\node[stage, above right=1.5cm and 2cm of ready] (e3) {Ethical\\Governance};

\node[stage, below right=1.5cm and 2cm of ready] (e4) {Institutional\\Legitimacy};

% Bottom
\node[outcome, below=3.2cm of ready] (out) {Sustainable \& Equitable AI-Enabled Disease Surveillance};

% Arrows
\draw[arr] (cap) -- (ready);
\draw[arr] (e1) -- (ready);
\draw[arr] (e2) -- (ready);
\draw[arr] (e3) -- (ready);
\draw[arr] (e4) -- (ready);
\draw[arr] (ready) -- (out);

\end{tikzpicture}

\caption{Conceptual framework illustrating AI implementation readiness as an emergent socio-technical condition. Technical capability enables surveillance, while epistemic adequacy, distributive justice, ethical governance, and institutional legitimacy collectively determine whether AI can be implemented responsibly and sustained within routine public health practice.}\label{fig:conceptual}
\end{figure}

Accordingly, this study conceptualizes implementation readiness as an emergent condition produced by the convergence of these four dimensions. AI capability provides the technical foundation for surveillance, but sustainable implementation becomes possible only when epistemic, ethical, distributive, and institutional conditions are simultaneously satisfied. The proposed framework therefore shifts analytical attention from evaluating predictive models toward understanding the broader conditions that enable responsible and equitable AI-enabled public health surveillance. Figure~\ref{fig:conceptual} illustrates how technical capability and the four socio-technical dimensions converge to produce implementation readiness and sustainable surveillance.

% ----------------------------------------------------------
% Figure 3: Conceptual framework (TikZ)
% ----------------------------------------------------------

% ==========================================================
% 7. RESEARCH GAP
% ==========================================================

\section{Research Gap}

The literature demonstrates substantial progress in AI-enabled outbreak prediction, disease classification, and epidemiological forecasting~\cite{santangelo2023machine,villanuevamiranda2025artificial}. However, the implementation of AI within routine public health systems remains conceptually underdeveloped.

First, existing research is predominantly \textbf{algorithm-centered}. Most studies evaluate predictive accuracy, sensitivity, specificity, and computational performance, while giving comparatively limited attention to the organizational and institutional conditions required for real-world implementation~\cite{santangelo2023machine,cozzolino2025surveillance}.

Second, the literature offers limited understanding of \textbf{implementation readiness in LMICs}. Although resource-constrained health systems are frequently identified as important contexts for AI adoption, implementation challenges are often discussed as isolated issues of infrastructure, workforce, or governance rather than as interconnected socio-technical conditions~\cite{ciecierskiholmes2022artificial,alganad2026deploying}.

Third, there is a significant \textbf{philosophical gap}. Current discussions rarely integrate epistemic questions concerning the reliability and representativeness of health data, distributive questions regarding equitable access to digital infrastructure, and ethical questions surrounding governance, legitimacy, and public trust into a unified framework of AI implementation~\cite{fricker2007epistemic,floridi2018ai}.

This study addresses these gaps by proposing a conceptual model of AI implementation readiness that synthesizes technological, epistemic, ethical, distributive, and institutional dimensions rather than introducing another predictive surveillance algorithm.

% ==========================================================
% 8. DISCUSSION
% ==========================================================

\section{Discussion}

The findings indicate that the principal challenge of AI-enabled disease surveillance is not the absence of predictive capability, but the absence of implementation readiness. Across the reviewed literature, successful deployment consistently depends upon the interaction between technological infrastructure, knowledge production, ethical governance, distributive equity, and institutional legitimacy rather than algorithmic performance alone.

The four implementation barriers identified in this study should therefore be understood as mutually reinforcing rather than independent. Fragmented health data limits the epistemic quality of AI-generated knowledge; unequal digital infrastructure restricts the equitable distribution of surveillance capabilities; weak governance undermines the ethical legitimacy of health-data practices; and limited institutional readiness reduces both clinician adoption and public trust. Together, these conditions explain why technically capable AI systems frequently fail to become sustainable components of routine public health practice.

The proposed conceptual framework also has practical implications. For policymakers, it suggests that investments in interoperable health information systems, digital governance, and institutional capacity may be as important as investments in predictive technologies. For researchers, the framework provides an interdisciplinary foundation for examining AI readiness through empirical studies that combine technical, organizational, and philosophical perspectives.

% ==========================================================
% 9. LIMITATIONS
% ==========================================================

\section{Limitations}

This study is conceptual and relies exclusively on secondary literature; consequently, the proposed framework represents an interpretive synthesis rather than an empirically validated implementation model. The relationships between the identified dimensions are theoretically grounded but have not been tested through primary data collected within a specific healthcare system.

A second limitation concerns contextual diversity. Although the analysis focuses on low- and middle-income countries, LMIC health systems differ considerably in their digital infrastructure, governance capacity, and institutional arrangements. The framework should therefore be understood as a transferable conceptual model rather than a universally predictive one.

Future research should validate the framework through mixed-method investigations, including interviews with public health professionals, surveys of institutional readiness, comparative case studies across LMICs, and quantitative assessments of how epistemic, ethical, infrastructural, and institutional factors jointly influence AI implementation.

% ==========================================================
% 10. CONCLUSION
% ==========================================================

\section{Conclusion}

Artificial intelligence has considerable potential to strengthen public health disease surveillance through earlier outbreak detection, epidemiological forecasting, and more informed resource allocation~\cite{santangelo2023machine,wikmanjorgensen2024hospitalization}. However, this study argues that technological capability alone is insufficient to ensure successful implementation within resource-constrained health systems.

Using a qualitative problem discovery approach, the study identified four interconnected dimensions of AI implementation readiness: epistemic adequacy, distributive justice, ethical governance, and institutional legitimacy. These dimensions explain how fragmented health data, digital infrastructure inequality, governance challenges, and limited public trust collectively constrain the translation of AI innovation into routine public health practice.

The principal contribution of this paper is therefore conceptual rather than algorithmic. By integrating public health, artificial intelligence, epistemology, ethics, and digital governance into a unified framework, the study reframes AI implementation as a socio-technical and normative challenge. This framework provides a foundation for future empirical research and may support the development of more responsible, equitable, and sustainable AI-enabled disease surveillance in low-resource public health systems.

\bibliographystyle{unsrtnat}
\bibliography{ref}

\end{document}